\documentclass[a4paper]{PoS}
\usepackage{multirow}
\usepackage{subfigure}
\usepackage{verbatim}

\title{Recovery Time Measurements of Silicon Photomultipliers Using a Pulsed Laser}

\ShortTitle{Recovery Time Measurements of Silicon Photomultipliers Using a Pulsed Laser}

\author{\speaker{L. Gruber}$^{a}$, S. E. Brunner$^{a}$, C. Curceanu$^{b}$, J. Marton$^{a}$, A. Romero Vidal$^{c}$, A. Scordo$^{b}$, K. Suzuki$^{a}$, O. Vazquez Doce$^{d}$\\
\\
\llap{$^{a}$}Stefan Meyer Institute for Subatomic Physics, Austrian Academy of Sciences, Vienna, Austria\\
\llap{$^{b}$}INFN, Laboratori Nazionali di Frascati, Frascati (Roma), Italy\\
\llap{$^{c}$}Universidade de Santiago de Compostela, Santiago de Compostela, Spain\\
\llap{$^{d}$}Excellence Cluster Universe, Technische Universit\"at M\"unchen (TUM), Munich, Germany\\
\\        
E-mail: \email{lukas.gruber@oeaw.ac.at}}

\abstract{We performed an experimental study to determine the pixel recovery time of various Multi Pixel Photon Counters (MPPCs) in order to characterize their rate capability and double-hit resolution. The recovery time constant and its dependence on the operating voltage has been evaluated by measuring the photosensor response to two consecutive laser pulses with varying relative time differences of a few ns (2 -- 3\,ns) up to some 100\,ns using a waveform analysis technique. A Monte Carlo simulation tool is being developed to model the MPPC recovery process and interpret experimental data. In this context, the influence of after-pulsing, cross-talk and dark-noise on the recovery process can be studied.}

\FullConference{The European Physical Society Conference on High Energy Physics\\
                 22-29 July 2015\\
                 Vienna, Austria}

\begin{document}

\section{Introduction}

The Silicon Photomultiplier (SiPM) is a semiconductor photodetector which consists of multiple pixels of Avalanche Photodiodes working in Geiger-mode (G-APD). The SiPM response is determined by numerous effects such as after-pulsing, cross-talk, dark-noise or the pixel recovery time, the latter being directly related to the performance of the photodetector in high rate environments present in modern particle physics experiments.

In order to characterize the SiPM rate capability and double-hit resolution we performed an experimental study to determine the pixel recovery time for various SiPMs manufactured by Hamamatsu Photonics K.K. (HPK), which are called Multi Pixel Photon Counters (MPPCs). We tested three MPPCs with $25\times25\,\mathrm{\mu m^{2}}$, $50\times50$\,$\mu$m$^{2}$ and $100\times100$\,$\mu$m$^{2}$ pixel size (S10362-11-025U, -050U and -100U) and evaluated the recovery time and its dependence on the operating voltage. All sensors have a sensitive area of 1$\times$1\,mm$^{2}$. The main parameters of all tested devices are summarized in Table~\ref{tab:SiPMparameters}.

The MPPC is also the most promising device for photon detection in the AMADEUS trigger system~\cite{triggersystem}. The AMADEUS experiment~\cite{amadeus} is planned to be installed inside the KLOE detector at the DA$\Phi$NE collider of LNF-INFN~\cite{dafne}. Constraints such as limited space and the need to operate the detectors in a magnetic field favor the use of SiPMs. The detector system is located very close to the beam pipe and will be exposed to a very high rate shortly after the e$^{+}$/e$^{-}$ injection. Therefore, an overall rate capability is required. A charged kaon decay-mode tagging to distinguish K$^{+}$ and K$^{-}$ is crucial to suppress the background, as done for $\pi^{+}$/$\pi^{-}$ separation described in Ref.~\cite{pienu}. In this context, a good double-hit resolution is essential. The determination of the recovery time can give information about the applicability of the MPPC for this experiment.

Measurements of the MPPC recovery time are also recorded in Ref.~\cite{du} and Ref.~\cite{oide}, where the recovery time constant has been extracted by studying after-pulsing and dark-noise, i.e. by recording the amplitude of the after-pulse and the corresponding time it occurs after the initial breakdown. As reported, the method is limited to time differences between after-pulse and initial pulse >\,10\,ns. This standard method is in contrast to the one presented here, which represents a direct measurement of the recovery time by using two real incident light pulses with a defined time difference (down to 2\,ns) to determine the recovery curve and is therefore a more realistic situation.

\section{Experimental Technique and Setup}

The basic building block of the G-APD is a p-n junction operated in reverse-bias mode at a certain voltage $\mathrm{V_{BIAS}}$ (typically a few Volts) above the breakdown voltage $\mathrm{V_{BR}}$ in series with a quenching resistor ($\mathrm{R_{Q}}$) to terminate the self-sustained avalanche. The difference between bias voltage and breakdown voltage is usually referred to as over-voltage $\mathrm{V_{OVER} = V_{BIAS} - V_{BR}}$.

\begin{table*}[t]
  \centering
  \footnotesize
  \begin{tabular}{| l | c | c | c |}
  \hline
  \multicolumn{1}{|l|}{\multirow{2}{*}{Parameter}} & \multicolumn{3}{c|}{HPK MPPC S10362-11} \\
			     & -100U & -050U & -025U \\ \hline \hline
  Active area [mm$^{2}$] & 1$\times$1 & 1$\times$1 & 1$\times$1 \\ 
  Number of pixels & 100 & 400 & 1600 \\ 
  Pixel size [$\mu$m$^{2}$] & 100 $\times$ 100 & 50 $\times$ 50 & 25 $\times$ 25 \\ 
  Terminal capacitance [pF] & 35 & 35 & 35 \\ 
  Quenching resistor [k$\Omega$] & 85 & 105 & 200 \\ 
  Breakdown voltage [V] & 69.45 & 68.65 & 68.85 \\ \hline
  \end{tabular}
  \caption{Main SiPM parameters. The breakdown voltage $\mathrm{V_{BR}}$ has been measured. Other parameters are taken from the data sheets. The capacitance values are terminal capacitances of the device. Pixel capacitances can be evaluated using the total number of pixels (ignoring parasitic capacitances).}
  \label{tab:SiPMparameters}
\end{table*} 

The creation of an electron-hole pair results in a discharge of the junction capacitance $\mathrm{C_{J}}$ and a drop of the voltage across the junction with a characteristic time constant $\tau_{1}$ and a subsequent recovery with a time constant $\mathrm{\tau_{2} = R_{Q} \times C_{J}}$. Since $\tau_{2}$ is usually much larger than $\tau_{1}$ the output current pulse has the typical asymmetric shape and the time constant $\tau_{2}$ can be used to estimate the pixel recovery time $\mathrm{\tau_{R}}$ of the device ($\mathrm{\tau_{R}\,\sim\,\tau_{2}}$). This is however only an approximation since the values for the capacitances and resistors can vary and are depending on the operating conditions such as temperature and voltage. In addition, noise effects lead to deviations from the ideal pulse shape. Typical values are $\mathrm{R_{Q}\,\sim\,150\,k\Omega}$ and $\mathrm{C_{J}\,\sim\,20 - 350\,fF}$, resulting in an expected recovery time constant in the range of some ns up to a few 10\,ns depending on the pixel size. 

The recovery time constant can be evaluated by studying the MPPC response to two consecutive laser pulses with a varying relative time difference of a few ns (2 -- 3\,ns) up to some 100\,ns (1 -- 2\,$\mu$s). Using such a technique there are in general two different ways to determine the recovery time. Illuminating only one certain pixel of the photosensor to determine the recovery time of this single pixel or illumination of all pixels to evaluate the recovery time of the whole sensor, i.e. the average of the recovery times of all pixels. Focusing the laser beam onto a single pixel is rather difficult and therefore the second method was chosen. Therefore the light intensity of each pulse must be high enough to illuminate all the pixels of the sensor and the pulse width should be small compared to the MPPC recovery time. The intensity was estimated before data taking by using a PIN photodiode to ensure full illumination. 

The measurements were performed using a blue (404\,nm) pulsed laser with 32\,ps pulse width (FWHM). The signals of the photosensor were not amplified and were recorded directly with a waveform digitizer from CAEN with 5\,GHz sampling rate. Because of the large MPPC signals due to full illumination it was not needed to use a pre-amplifier, which anyway has a limited dynamic range. As a consequence, the impact of the electronics on the measurement results was reduced to a minimum. All measurements were done at room temperature ($\sim\,25\,^{\circ}\mathrm{C}$). In order to compensate for small temperature variations during measurements the temperature dependence of $\mathrm{V_{BR}}$ has been considered. At the AMADEUS experiment the MPPCs will also be operated at room temperature.

\begin{figure}[t]
  \centering
  \includegraphics[width=0.62\textwidth]{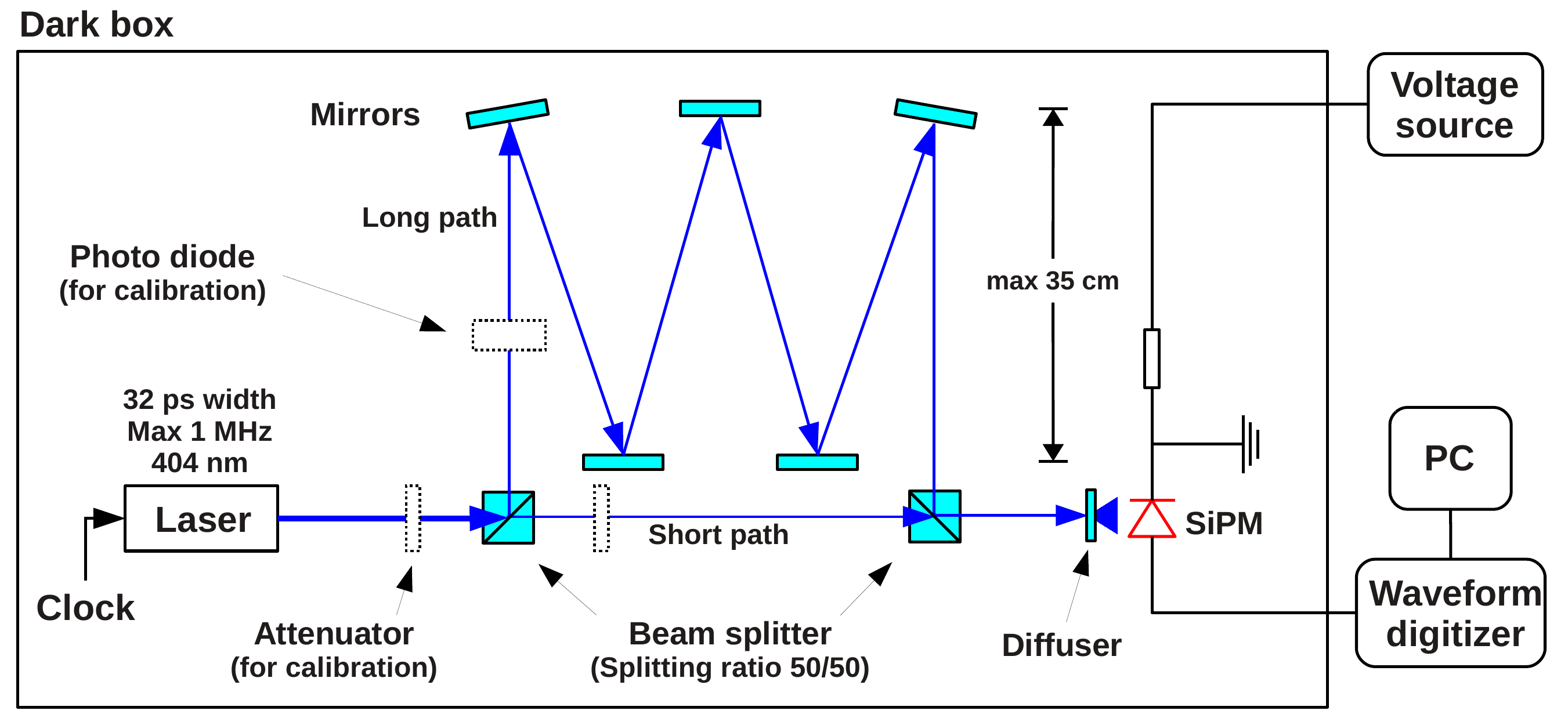}
  \includegraphics[width=0.62\textwidth]{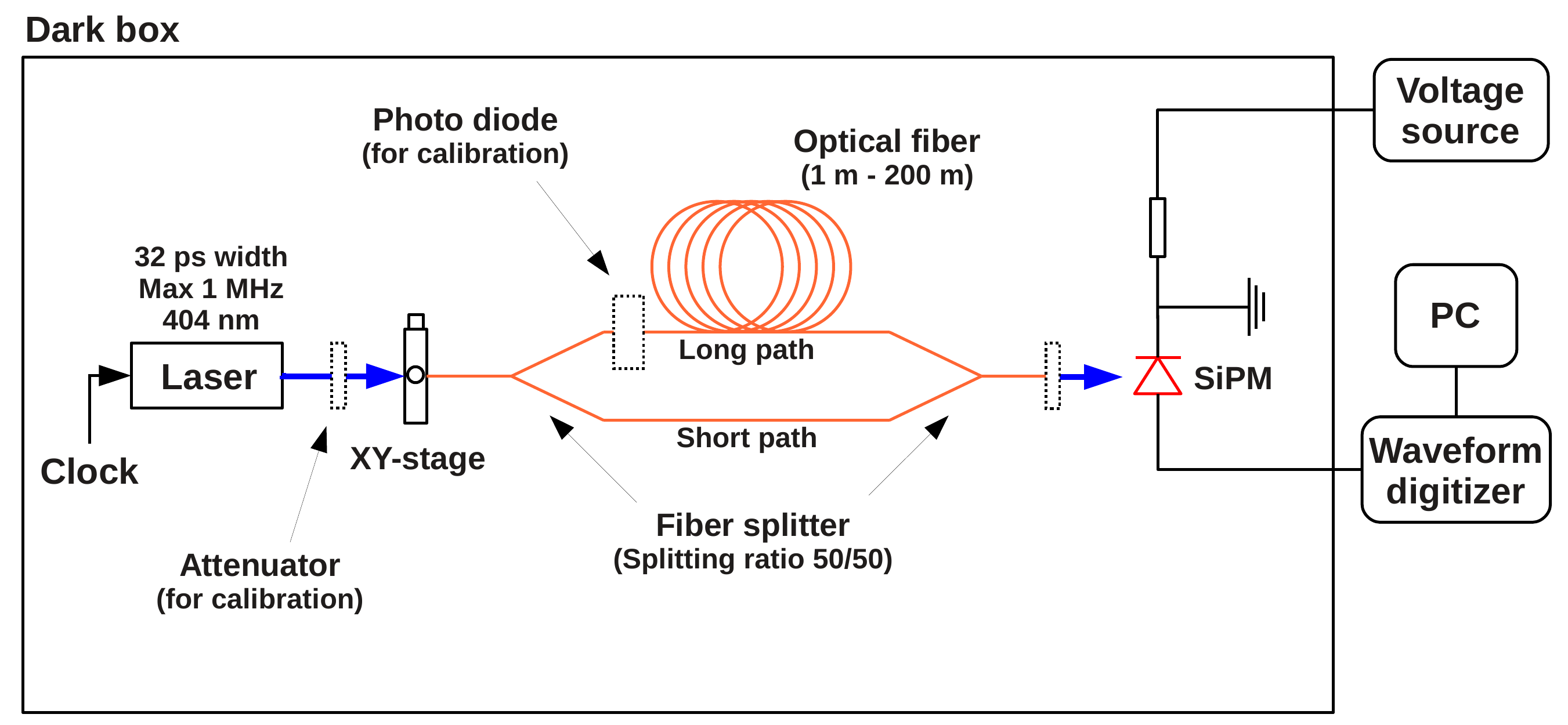}
  \caption{Schematic drawings of the different setups using mirrors (top) and optical fibers (bottom). }
  \label{fig:setup}
\end{figure}

The two pulses were generated by splitting the laser signal (short and long path), delaying one of the signals and merging them again. The delay of the second pulse was either done with multiple reflections using mirrors (up to 5\,ns) or for larger delays (up to 1\,$\mu$s) by coupling the light into optical fibers of different length (1\,m to 200\,m). Even larger time intervals between the two pulses could be realized without delay but simply by tuning the repetition frequency of the laser (10\,Hz to 1\,MHz). Using these three techniques we were able to cover a wide delay time range. However, the most interesting delay time region to measure is below 1\,$\mu$s, so above the maximum repetition rate of the laser. Therefore, most of the measurements were done using mirrors or fibers, respectively, while fixing the laser frequency to 1\,kHz. The mirrors were especially needed to provide short delay times to investigate the double-hit resolution and its limits. The measurement setups are shown schematically in Fig.~\ref{fig:setup}.

\section{Data Analysis and Interpretation}
In order to extract the recovery time constant the data were interpreted by performing a full waveform analysis. The procedure to extract the recovery time from the measured recovery curve, i.e. the relation between fraction of recovery and time difference (delay) of two consecutive light pulses, for a specific MPPC and fixed over-voltage is explained in the following. 

\begin{figure}[t]
  \centering
  \includegraphics[width=0.45\textwidth]{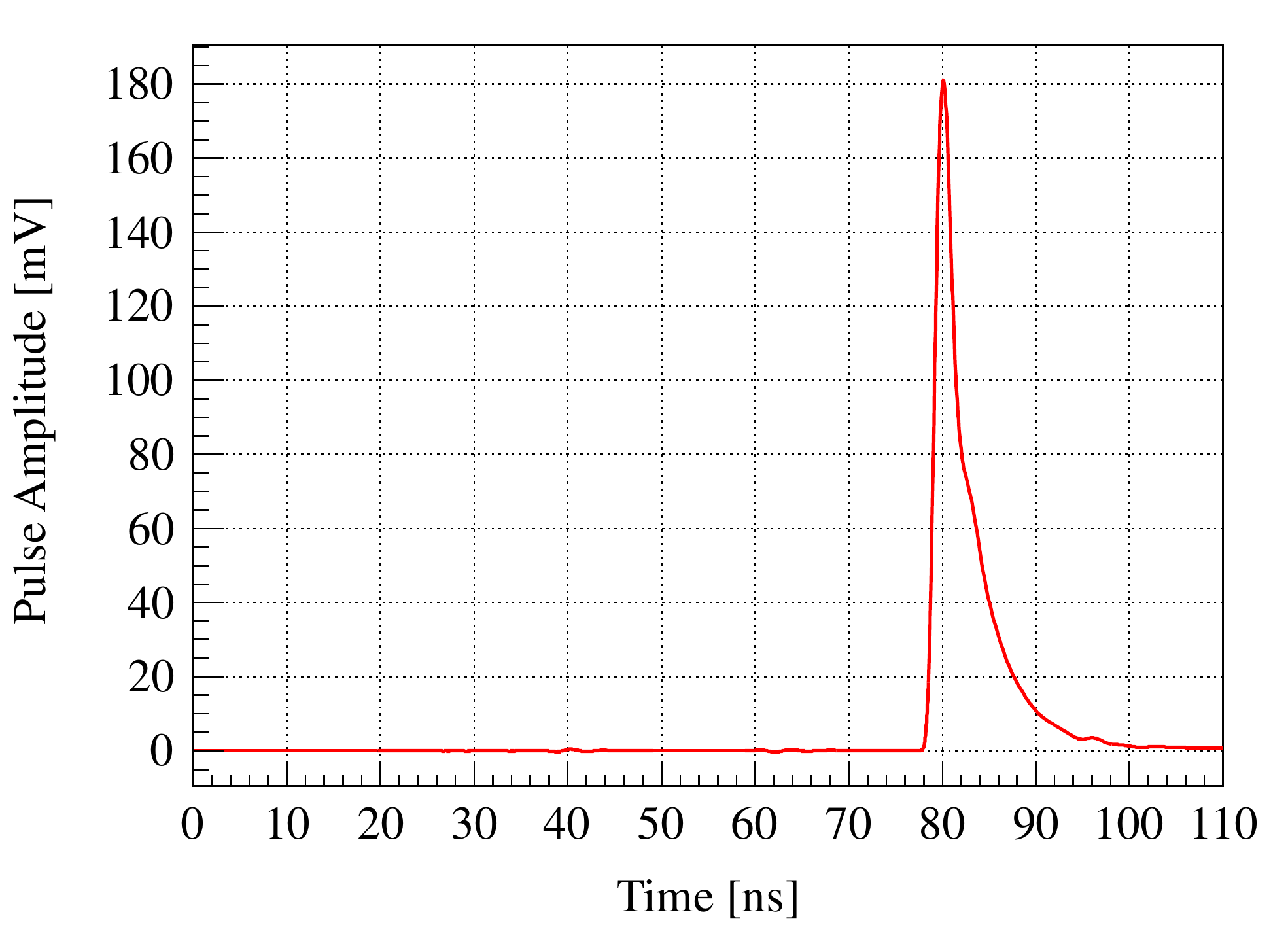}
  \caption{Template waveform obtained by measuring the signal when the long path is blocked. The figure shows an average of 1000 waveforms.}
  \label{fig:template}
\end{figure}

\begin{figure}[t]
  \centering
  \subfigure{
    \includegraphics[width=0.47\textwidth]{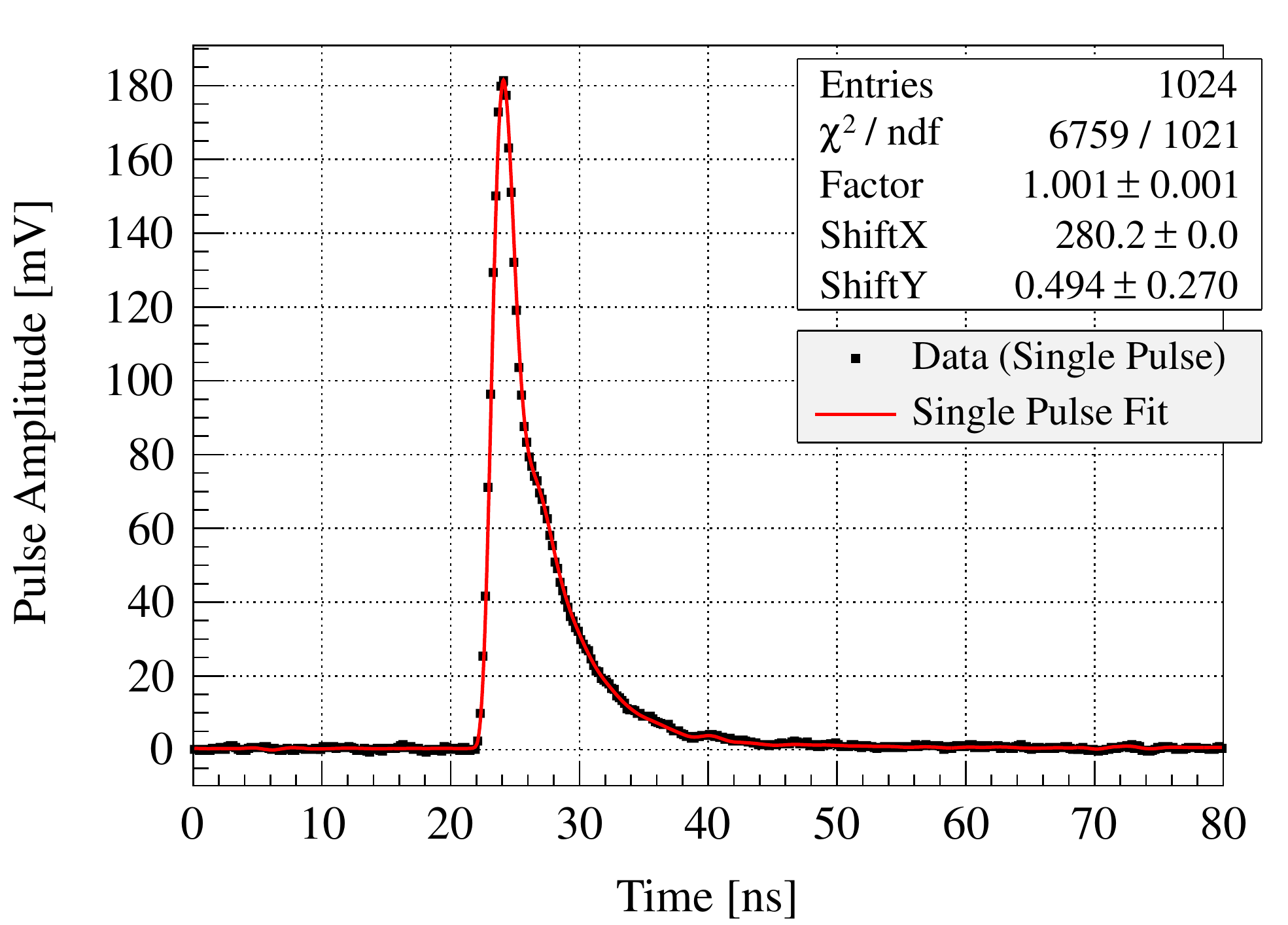}
  }
  \subfigure{
    \includegraphics[width=0.471\textwidth]{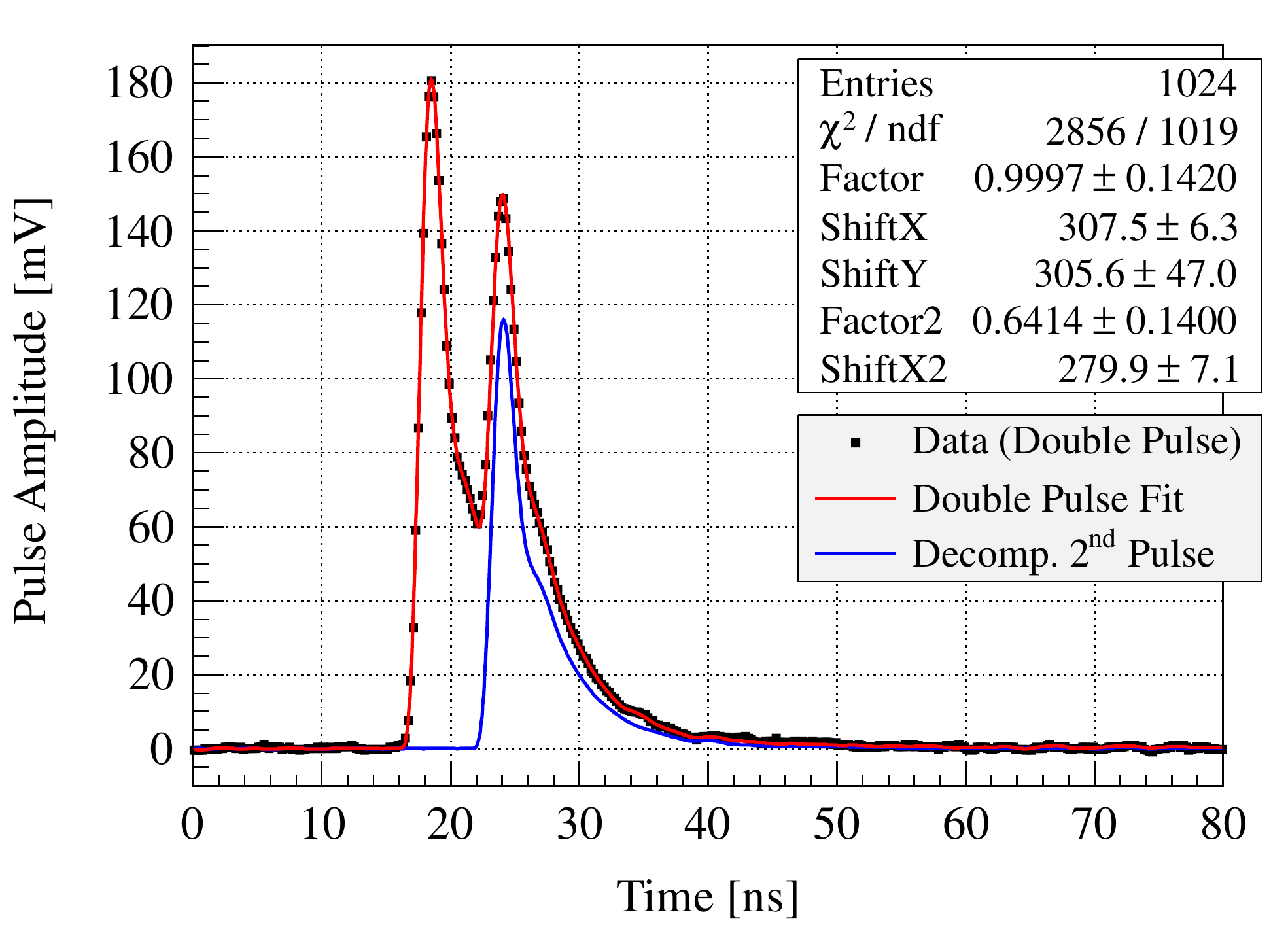}
  }
  \caption{Example waveforms of a measurement with an MPPC from Hamamatsu with $25\times25\,\mathrm{\mu m^{2}}$ pixel size operated at $\mathrm{V_{OVER} = 1\,V}$. The figure on the left shows the signal when the short path is blocked (single pulse), the plot on the right indicates the signal when both paths are connected. In each case the signals are fitted using a template function, which is scaled by a factor and shifted horizontally and vertically.}
  \label{fig:wavefit}
\end{figure}

Firstly a template waveform used for fitting the data is created. Therefore the second (long) path is disconnected (fibers) or blocked (mirrors), respectively, resulting in a single output pulse as indicated in Fig.~\ref{fig:template}. Then the long path is reconnected or unblocked while the short path gets interrupted. In the following, each single event is fitted by shifting (vertically and horizontally) and scaling the template waveform obtained before (see Fig.~\ref{fig:wavefit}), resulting in three fit parameters. Finally the short path is reconnected and the obtained double pulse is fitted by adding a second template function properly scaled and shifted, which results in two additional parameters. In the example shown in Fig.~\ref{fig:wavefit}, the second pulse is delayed by 5.5\,ns using a fiber of 1\,m length. Steps two and three are applied to 1000 events using the same template function to obtain average values and corresponding errors of pulse amplitudes and delay time. In the next step, the recovery fraction is estimated by determining the reduction of the second pulse, which is estimated by comparing the amplitude of the single pulse when the first (short) path is blocked and the amplitude of the decomposed second pulse. The procedure is then repeated for different delay lengths in order to obtain the recovery curve and finally extract the recovery time constant of the MPPC. The template function has to be redefined only when changing the photosensor or the over-voltage.

\begin{figure}[t]
  \centering
  \includegraphics[width=0.32\textwidth]{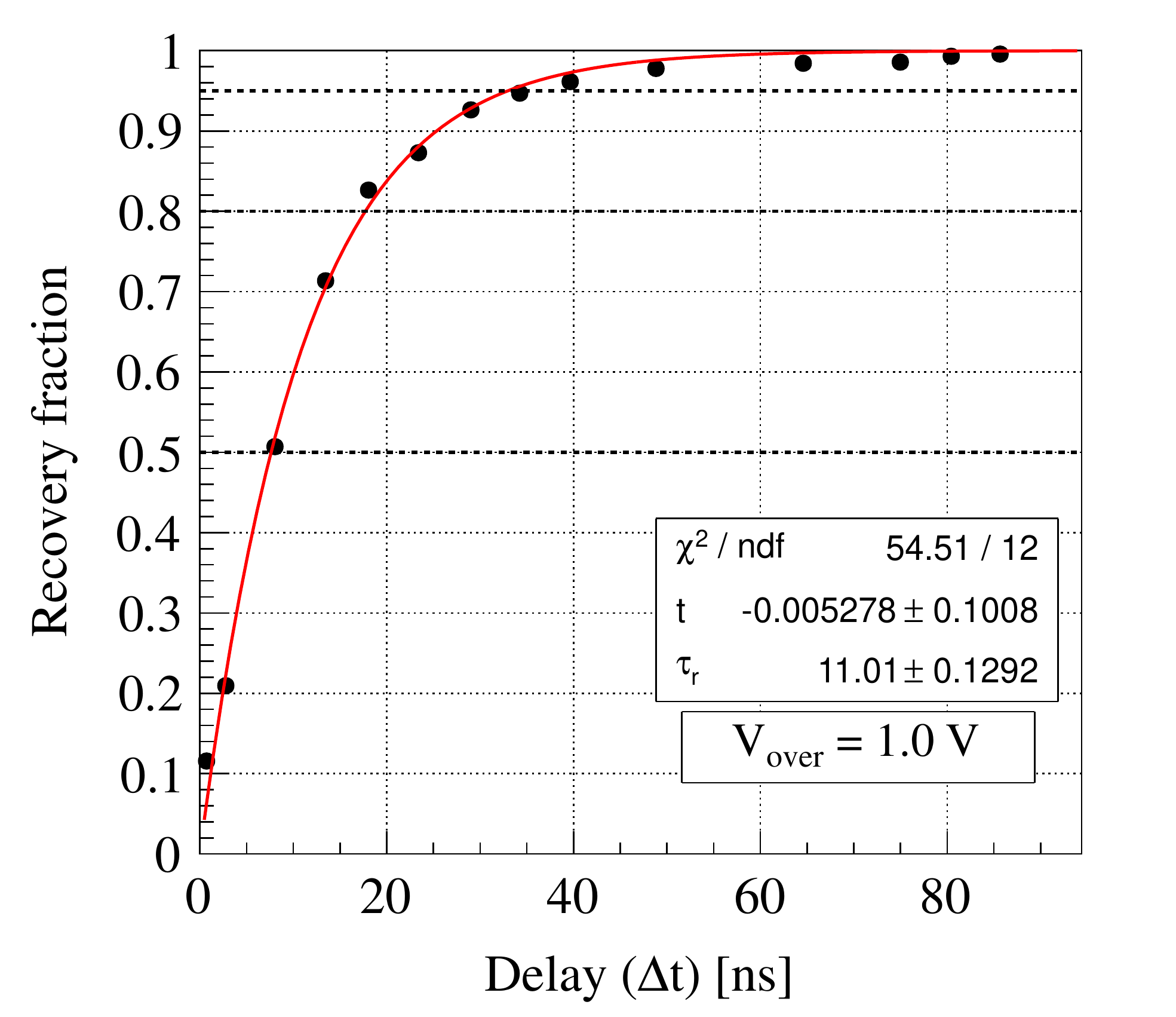}  
  \includegraphics[width=0.32\textwidth]{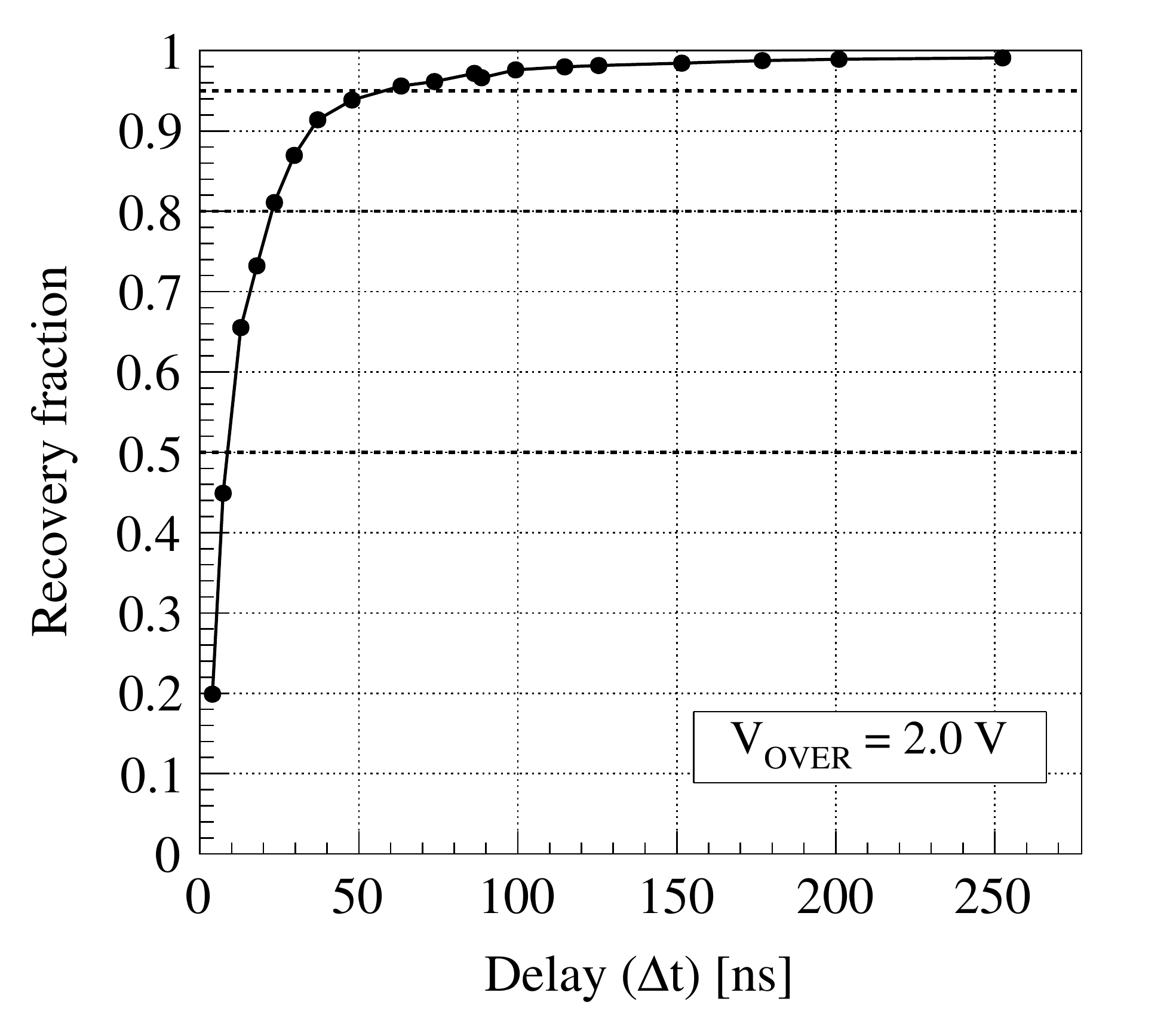} 
  \includegraphics[width=0.32\textwidth]{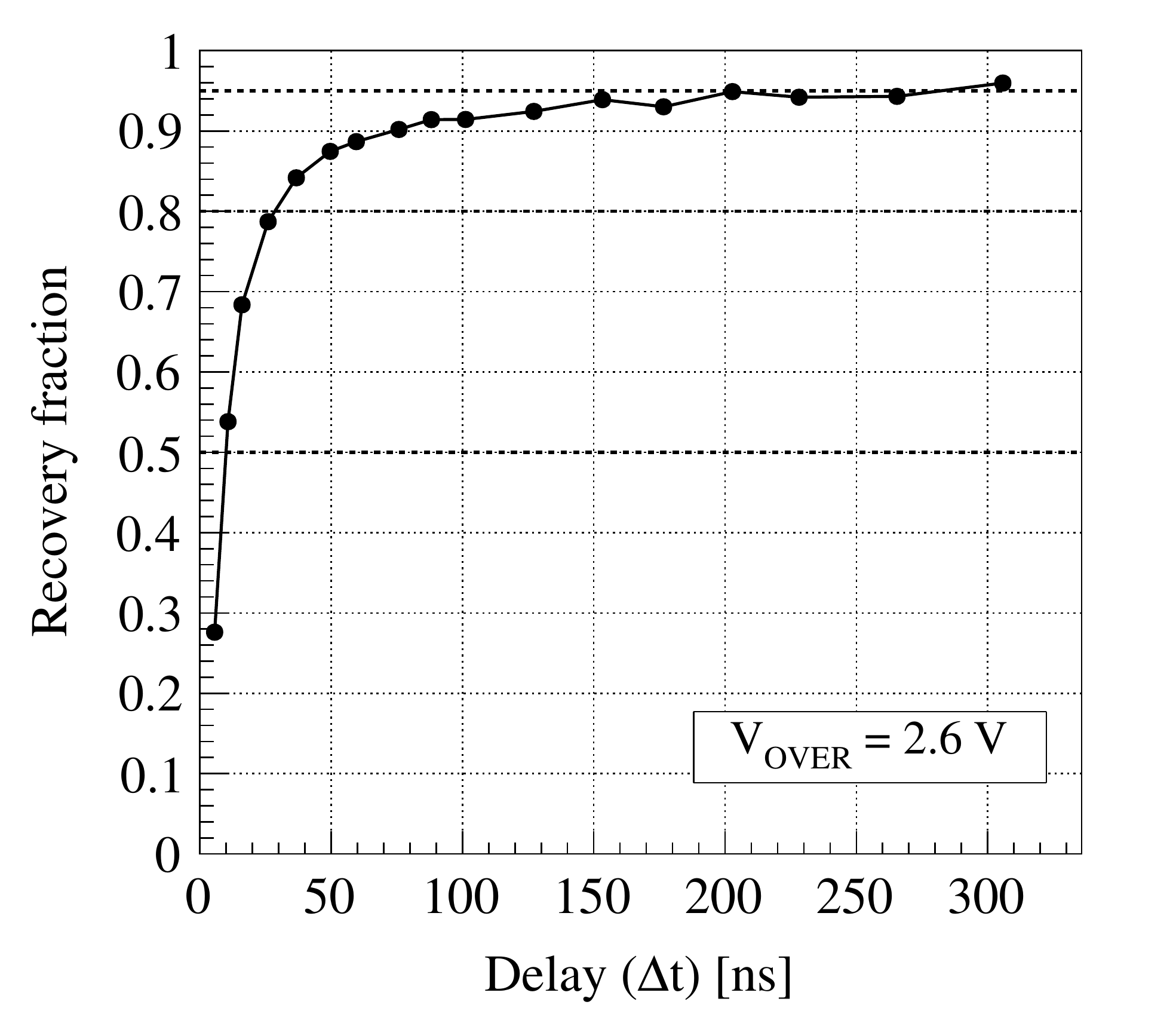}  
  \caption{Recovery curves for S10362-11-050U MPPCs operated at different over-voltages. The black dots are the data points. The solid black lines are added to guide the eye. Recovery fractions of 50\,\%, 80\,\% and 95\,\% are indicated by the horizontal dashed lines. For low operating voltages the recovery curve is fitted with a simple function shown in red.}
  \label{fig:recoverycurves}
\end{figure}

The recovery curves of the MPPC with 50\,$\mu$m$^{2}$ pixel size are shown in Fig.~\ref{fig:recoverycurves} for different operating voltages. For low bias voltages ($\mathrm{V_{OVER} \leq 1\,V}$) the influence of noise effects is very small and the recovery curves can be fitted with a simple function of the form
\begin{equation}
  y(\Delta t) = 1 - \exp \left[-(\Delta t-t)/\tau_{\mathrm{R}}\right],
\end{equation}
which describes the recovery with the time constant $\mathrm{\tau_{R}}$. For a larger gain the MPPC response is influenced by a multitude of effects and thus the analytical description of the recovery process is rather difficult. However, from the measured curves we can still extract the time needed to achieve a certain fraction of recovery, as indicated in the plots.

\section{Results and Discussion}
\subsection{Recovery Time and Rate Capability}
The results of the measurements are summarized in Table~\ref{tab:recoveryconstants} for the three tested MPPCs. The measured time constants are compared with the simple approximation $\mathrm{\tau_{R} \sim R_{Q} \times C_{J}}$. The obtained values for the RC time constants roughly agree with the measured recovery times. It is interesting to notice that the recovery is substantially prolongated when increasing the bias voltages, which is due to the rapid increase of the cross-talk and after-pulse probability, respectively, as well as dark count rate, as reported in Ref.~\cite{eckert}. At moderate operating voltages ($\mathrm{V_{OVER}\,\leq\,1\,V}$) the times needed for full recovery (100\,\%) are in good agreement with the values given by Hamamatsu. We obtain values of about 200\,ns for the 100U MPPC device, 50\,ns in case of the 050U sensor type and 20\,ns for the 025U pixels. In the data sheets values of 100\,--\,200\,ns, 50\,ns and 20\,ns are stated for the respective devices. Thus the maximum rate capability can be estimated to about 5\,MHz, 20\,MHz and 50\,MHz for 100U, 050U and 025U devices operated at moderate bias. It has to be noted that one should confirm in a separate measurement if the MPPCs can really be operated continuously at such high rates. However, the values suggest that especially the small pixel devices (small $\mathrm{C_{J}}$) can stand substantially higher rates than standard PMTs, which can be operated usually up to rates in the order of 1\,MHz.

\begin{table}[t]
  \centering
  \footnotesize
  \begin{tabular}{| l | c  c | c  c  c | c  c  c |}
  \hline
  MPPC type 						& \multicolumn{2}{c|}{S10362-11-100U} & \multicolumn{3}{c|}{S10362-11-050U} & \multicolumn{3}{c|}{S10362-11-025U} \\ \hline \hline
  Over-voltage (V$_{\mathrm{OVER}}$)                	& +\,1.0\,V   & +\,1.7\,V & 	+\,1.0\,V 	& +\,2.0\,V & +\,2.6\,V  & +\,1.0\,V  & +\,2.0\,V 	& +\,4.3\,V \\
  $\mathrm{\Delta t}$ for 50\,\% recovery [ns]       	& 30  & 50	 & 8 & 10 	& 12 & 2.5 & 3 	& 3 \\ 
  $\mathrm{\Delta t}$ for 80\,\% recovery [ns]       	& 75  & 140 & 18 & 22 & 29 & 5.5 & 7 & 10 \\ 
  $\mathrm{\Delta t}$ for 95\,\% recovery [ns]       	& 140 & 800 & 33 & 60 & 280 & 11 & 13 & 130  \\ 
  Recovery time constant ($\mathrm{\tau_{R}}$) [ns] 	& 46.0  & -- & 11.0 & --  & -- & 3.5       & --        	& -- 	 	 \\ 
  RC time constant [ns]                             	& 29.8  & -- & 9.2 & -- & -- & 4.4 	    & -- 		& -- 	 	 \\  \hline 
  \end{tabular}
  \caption{Recovery time constants for the tested MPPCs operated at different bias voltages. The time differences between two light pulses ($\mathrm{\Delta t}$) needed to achieve a certain fraction of recovery have been extracted from the recovery curves. The error on $\mathrm{\Delta t}$ values is about 10\,\%. The recovery time constant ($\mathrm{\tau_{R}}$) can be obtained by fitting the recovery curves measured at low bias voltage, i.e. low gain.}
 \label{tab:recoveryconstants}
\end{table} 

\subsection{Double-hit Resolution}
\begin{figure}[t]
  \centering
  \includegraphics[width=.36\textwidth,angle=-90]{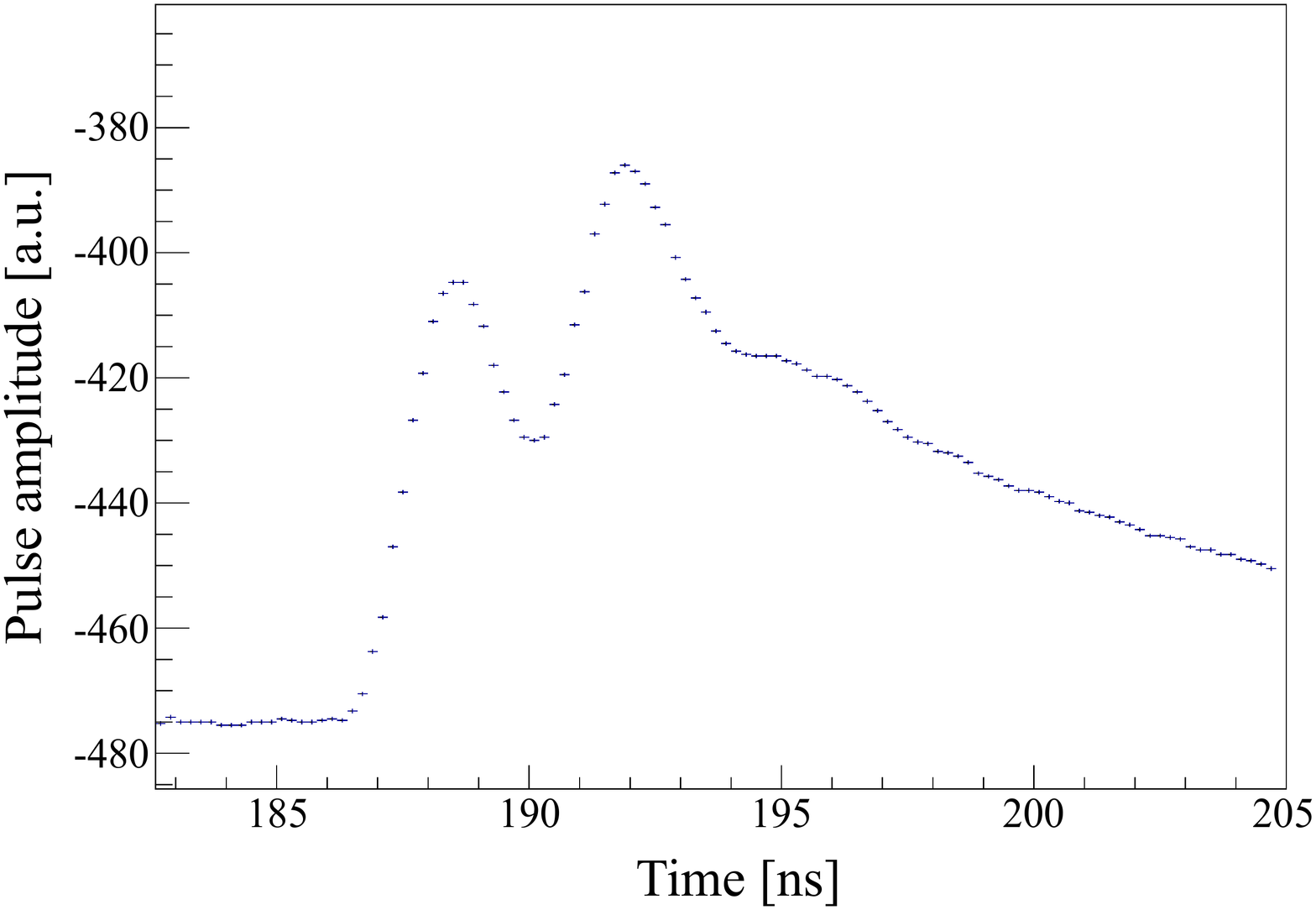}
  \caption{Illustration of double pulse with about 3.5\,ns delay measured with 050U device.}
  \label{fig:doublepulse}
\end{figure}

Using a technique of reading out the MPPC signals with a fast waveform digitizer we could resolve two adjacent light pulses down to about 2 ns and 3 ns delay between pulses with the 025U and 050U devices, respectively, which can be seen as the limit of the double-hit resolution of these sensors. In this regime the fraction of recovered pixels is about 30\,\%. Fig.~\ref{fig:doublepulse} shows a double pulse measured with the 050U sensor type operated at 2\,V over-voltage and illustrates that the two pulses with a time difference of about 3.5\,ns can be clearly separated. In general one can say that small pixel sizes (small $\mathrm{C_{J}}$) lead to fast recovery and short pulse width and therefore improved double-hit resolution.

\section{Conclusion and Outlook}
The recovery time of Hamamatsu MPPCs has been determined by measuring the sensor response to two consecutive laser pulses and applying a waveform analysis technique. In this context, the maximum rate capability and double-hit resolution of the photodetectors could be estimated. It was shown that small pixel devices with $25\times25\,\mathrm{\mu m^{2}}$ pixel size operated at moderate bias could potentially stand rates up to 50\,MHz and could be used to identify signals separated by only 2\,ns. For large pixels it was found that these values degrade by about a factor of 10. For higher operating voltages ($\mathrm{V_{OVER}\,>\,1\,V}$) the MPPC recovery is substantially prolongated due to the increasing influence of cross-talk, after-pulsing and dark-noise, resulting in times needed for full recovery up to $2\,\mathrm{\mu s}$.

In order to better describe the SiPM recovery process a Monte Carlo simulation tool is being developed. The simulation is going to be compared to experimental data to extract the recovery time constants and to study the influence of the individual parameters determining the SiPM response. In the simulation, the bias voltage (gain) recovery is described by an exponential recovery with the time constant $\mathrm{\tau_{R}}$. After-pulsing is characterized by an after-pulse probability $\mathrm{P_{AP}}$ and a time constant $\mathrm{\tau_{AP}}$. Cross-talk can be described using a cross-talk probability $\mathrm{P_{CT}}$ and the dark-noise is given by the dark-count rate DCR. All the simulation parameters are determined by fitting the data points. First results are in good agreement with experimental results and show almost no dependence of the recovery time constant on the operating voltage, as also reported in Ref.~\cite{oide}.


\begin{thebibliography}{99}

\bibitem{triggersystem}
M. Bazzi et al., \emph{Experimental tests of the trigger prototype for the AMADEUS experiment based on Sci-Fi read by MPPC}, Nucl. Instr. Meth. A 671 (2012) 125-128. 

\bibitem{amadeus}
J. Zmeskal et al., \emph{The AMADEUS experiment -- precision measurements of low-energy antikaon nucleus/nucleon interactions}, Nucl. Phys. A 835 (2010) 410.

\bibitem{dafne}
D. Alesini et al., \emph{DA$\Phi$NE upgrade for the SIDDHARTA run}, LNF-06-33-IR, 2006.

\bibitem{pienu}
K. Yamada et al., \emph{Pion Decay-Mode Tagging in a Plastic Scintillator Using COPPER 500-MHz FADC}, IEEE Trans. on Nucl. Sci., Vol. 54, Issue 4, pp. 1222-1226, 2007.

\bibitem{du}
Y. Du, F. Reti\`ere, \emph{After-pulsing and cross-talk in multi-pixel photon counters}, Nucl. Instr. Meth. A 596 (2008) 396-401.

\bibitem{oide}
H. Oide et al., \emph{Study of afterpulsing of MPPC with waveform analysis}, PoS(PD07)008 (2007).

\bibitem{eckert}
P. Eckert et al., \emph{Characterisation studies of Silicon Photomultipliers}, Nucl. Instr. Meth. A 620 (2010) 217-226.

\end{thebibliography}
\end{document}